\documentclass[preprint,aps,prb,showpacs,amsmath]{revtex4}
\usepackage{graphicx}
\usepackage{color}
\usepackage{amsmath}
\usepackage[american]{babel}

\begin{document}

\newcommand{\BaFeAs}{BaFe$_{\textnormal{2}}$As$_{\textnormal{2}}$\ }
\newcommand{\CaFeAs}{CaFe$_{\textnormal{2}}$As$_{\textnormal{2}}$\ }
\newcommand{\BaKFeAs}{Ba$_{\textnormal{1-x}}$K$_{\textnormal{x}}$Fe$_{\textnormal{2}}$As$_{\textnormal{2}}$\ }
\newcommand{\BaFeCoAs}{Ba(Fe$_{\textnormal{1-x}}$Co$_{\textnormal{x}}$)$_{\textnormal{2}}$As$_{\textnormal{2}}$\ }
\newcommand{\nBaFeCoAs}{Ba(Fe$_{\textnormal{0.9375}}$Co$_{\textnormal{0.0625}}$)$_{\textnormal{2}}$As$_{\textnormal{2}}$}
\newcommand{\LOFA}{LaOFeAs}
\newcommand{\tc}{$T_c$}
\newcommand{\half}{\frac{1}{2}}
\newcommand{\sixteenth}{$\frac{1}{16}$}

\title{Effects of cobalt doping and three-dimensionality in \BaFeAs}

\author{A.F. Kemper}
\email{kemper@qtp.ufl.edu}
\author{C. Cao}
\author{P.J. Hirschfeld}
\author{H-P. Cheng}
\affiliation{Department of Physics, University of Florida, Gainesville, FL 32611, USA}

\date{\today}

\begin{abstract}
We investigate the dual roles of a cobalt impurity in the Ba-122
ferropnictide superconductor in the state with coexisting
collinear spin density wave (SDW) order as a dopant and as a
scattering center, using first principles electronic structure
methods. The Co atom is found to dope the FeAs plane where it is
located with a single delocalized electron as expected, but also
induces a strong perturbation of the  SDW ground state of the
system.  This  in turn induces a stripe-like modulation of the
density of states in nearby planes which may be observable in STM
experiments. The defect is found to have an intermediate strength
nonmagnetic
 scattering potential with a range of roughly
1\ \AA, and the Co gives rise to a smaller but longer range
magnetic scattering potential.  The impurity potential in both
channels is highly anisotropic, reflecting the broken symmetry of
the SDW ground state.  We give values for the effective Co
potentials for each $d$ orbital on the impurity and nearby sites.
The calculation also shows a clear local resonance comprised of Co
states about 200meV above the Fermi level, in quantitative
agreement with a recent report from STM. Finally, we discuss the
issue of the effective dimensionality of the 122 materials, and
show that the hybridization of the out-of-phase As atoms leads to
a higher density of states between the FeAs planes relative to the
1111 counterparts.
\end{abstract}

\maketitle

\section{Introduction}
The recent discovery of superconductivity in the oxypnictide
systems {\textit{R}}OFeAs, where {\textit{R}} is one of the rare
earths La, Sm, Pr, Nd and Gd, has generated a great deal of
interest in  these and related compounds. Soon thereafter,
superconductivity was reported upon hole-doping of the oxygen-free
pnictides {\textit{R}}Fe$_{\textnormal{2}}$As$_{\textnormal{2}}$
\cite{m_rotter_08,n_ni_08a}, where {\textit{R}} in this case is
one of the alkaline earth metals Ca, Sr and Ba, formed in the
ThCr$_{\textnormal{2}}$Sr$_{\textnormal{2}}$ structure. In the
{\textit{R}}OFeAs-type materials (which we refer to as  ``1111")
critical temperatures of 56 K have been achieved,\cite{g_wu_08}
whereas in the
{\textit{R}}Fe$_{\textnormal{2}}$As$_{\textnormal{2}}$-type
materials (``122"), \tc ~has thus far been limited to 38
K.\cite{m_rotter_08} In the latter materials, the initial (hole)
doping was accomplished by substitution of the alkaline
earth metal by an alkali, such as K.\cite{m_rotter_08}\\

With the discovery of superconductivity at 22K upon cobalt-doping
of \BaFeAs by Sefat et al.,\cite{a_sefat_08} the oxygen-free
``122" pnictide materials have now been shown to also superconduct
upon electron doping.  Since single crystals of relatively high
quality can be grown without the difficulty of handling the alkali
dopant, these materials have become quite popular for
investigation of fundamental properties of the Fe-pnictides.
Furthermore, the Co atom substitutes directly for an Fe, in the
FeAs plane where it is believed that pairing and magnetism
originate. After doping the plane, the nominal electronic
configuration of Co is quite similar to the Fe it replaces, but
represents an effective scattering potential localized primarily
in the plane as well. One might expect that the scattering
potential sampled by quasiparticles in the FeAs plane due to a Co
would be considerably larger than an out-of-plane alkali dopant
due to its location in the plane; however this argument, familiar
from the cuprates, must be critically examined
if the systems are more three-dimensional in nature.\\

 Another possibly important difference between Co and the alkali dopant is
related to spin: cobalt frequently has a strong intrinsic magnetic
moment, and in its metallic form it orders ferromagnetically.  The
Fe-pnictide superconductors manifest a ``spin-density wave" (SDW)
ordering transition at high temperatures, and enhanced spin
fluctuations due to proximity to the SDW state have been suggested
as a pairing
mechanism.\cite{s_graser_08,k_kuroki_08,xl_qi_08,v_barzykin_08,y_bang_08,z_yao_09,r_sknepnek_09,f_wang_09,a_chubukov_08}.
Thus, one can expect that the introduction of an ion with a
different magnetic moment than Fe will have a strong effect on the
SDW state. Indeed, it is found that   Co-doping suppresses and
splits the magnetic and structural phase transitions, which are
degenerate in the 122 parent compounds\cite{canfield_09}. The
magnetic transition temperature, which is suppressed more rapidly,
 continues to decrease
as superconductivity arises; therefore a range of Co doping
corresponds to  a coexistence of the two
states\cite{s_takeshita_08,x_wang_08,j_chu_08,f_ning_08,n_ni_08,r_gordon_08},
which recent NMR experiments have shown to be homogeneous at the
atomic scale\cite{bobroff_09}.  To understand the destruction of
the SDW by doping, as well as the possibilities of coexistence of
the two states,  it will be interesting to investigate the effect
of the Co dopant on the SDW locally, and to ask to what extent Co
acts as a magnetic or nonmagnetic scatterer.

The Fe-pnictides are of great importance not only because of their
potentially high critical temperatures, but also because they
provide the possibility of comparing the even higher temperature
cuprate materials to another class of high-$T_c$ superconductors.
It is frequently speculated that superconductivity at high
temperatures arises in the cuprates because of their highly
layered nature which enhances electronic correlations. With the
discovery of the  1111-ferropnictides, DFT calculations indicated
that the  density of states near the Fermi level was localized in
the FeAs planes, and that the Fermi surfaces were nearly 2D,
weakly corrugated cylinders\cite{c_cao_08}.  The 122s are similar
in this respect, as reported by LDA studies\cite{a_sefat_08}, but
have a distinctly larger z-dispersion in some of the bands; and
Fermi surfaces near the M points appear to have significantly
larger corrugations.  In addition,  several studies have recently
reported an unexpectedly small anisotropy in c-axis versus a-axis
resistivity, penetration depth, and upper critical
field.\cite{m_tanatar_08,n_ni_08} To investigate further, Prozorov
et al.\cite{r_prozorov_09} compared the anisotropy in the London
penetration depth for \BaKFeAs, \BaFeCoAs and
NdFeAs(O$_{\textnormal{1-x}}$F$_{\textnormal{x}}$), and reported
low-temperature ratios ($\lambda_c/\lambda_a$) of approximately 6,
7 and 18, respectively. These data suggest that the iron-based 122
superconductors are quite complex compared to the cuprates, with
multi-sheeted Fermi surfaces, at least some of which are quite
3-dimensional.  It is not yet clear why these materials are more
three dimensional, if the 3D character is a strong function of
doping, or if this property is useful or harmful for
superconductivity.\\

 In this paper, we shall address the effect of
Co-doping in \BaFeAs using first-principles calculations. We find
that, in the SDW state, Co dopes the system with a single elecron
as expected, and that it breaks the symmetry between up and down
electrons. The resulting scattering potential is found to be of
order the effective bandwidth in the system, making it an
intermediate-strength potential scatterer, while the magnetic part
of the potential is considerably smaller, but quite long range.
The potential in both channels displays a strongly anisotropic
twofold character reflecting the collinear SDW.   Curiously,
although we find that the Co electron principally dopes the FeAs
plane where it is located, there are also significant effects on
the local density of states of neighboring planes, which may be
visible to STM. Localized Co resonances at -800 and +200 meV
should also be observable in STM. Finally, we discuss the origin
of the reported anisotropy of the 122s, as compared to the 1111s.
We find that the inter-plane local density of states near the
Fermi surface has a strong As p$_{\textnormal{z}}$ character. Due
to the different symmetry positions that the As atoms occupy in
the 1111s as compared to the 122s, there is a stronger overlap of
the As p$_{\textnormal{z}}$ orbitals in the 122s, which gives the
122s their comparatively stronger three-dimensional nature. Co
doping appears to suppress this hybridization.

\section{Calculational details and structure}

\begin{figure}[ht]
\includegraphics[height=3in, clip = true]{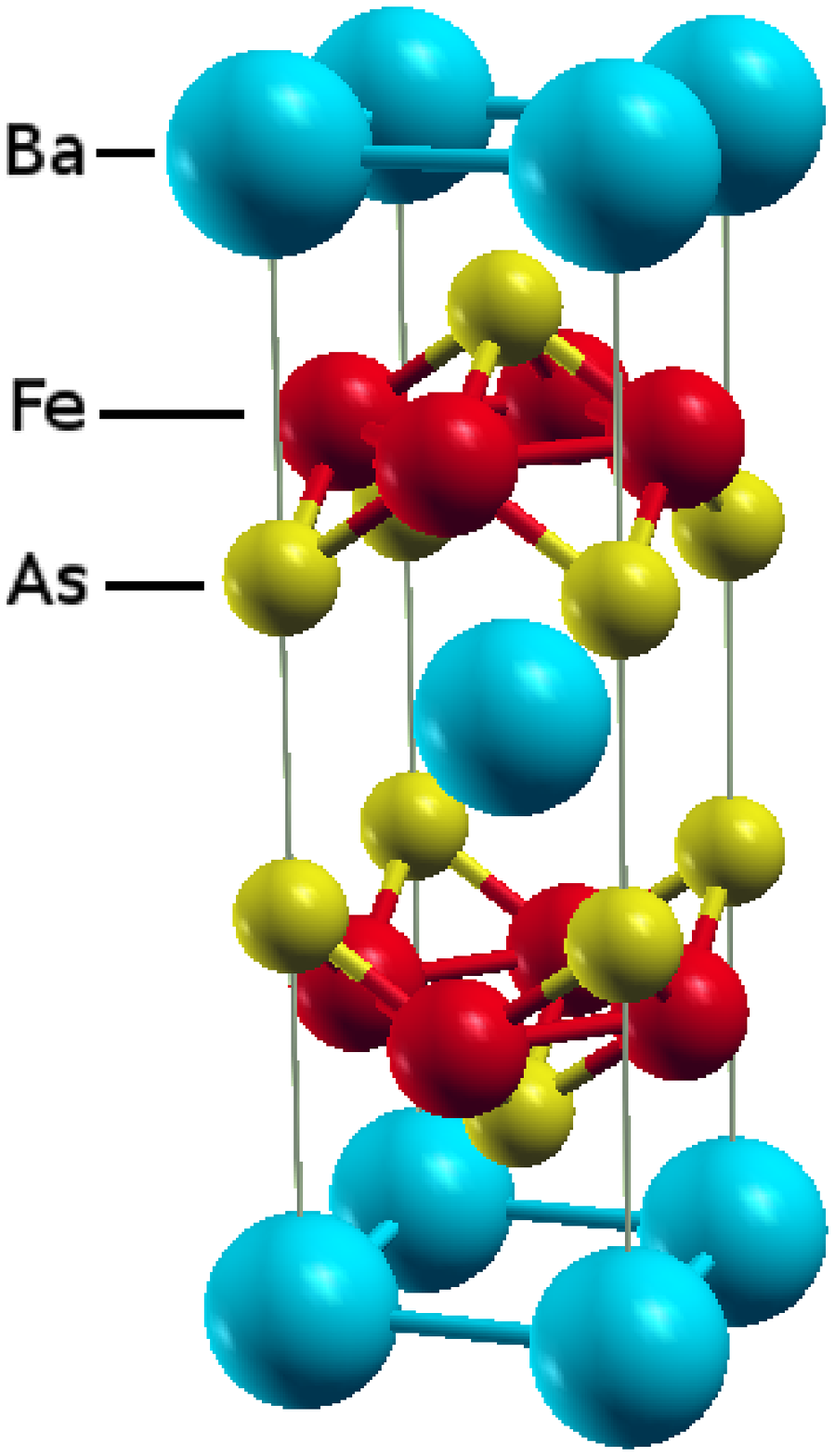}
\caption{(Color online) Structure of \BaFeAs.}
\label{fig:smallstructure}
\end{figure}

The structural relaxation calculations were performed using
Density Functional Theory (DFT)\cite{wkohn65,p_hohenberg_64} within the
generalized gradient approximations as implemented in the
Quantum-ESPRESSO\cite{pwscf} package, which uses a plane wave
basis. We used the Perdew-Burke-Ernzerhof\cite{jperdew96}
exchange-correlation functionals and ultrasoft pseudopotentials.
The use of ultrasoft pseudopotentials enabled us to utilize an energy
cut-off of 40 Ry for the plane wave basis, while the density
cut-off was taken to be 400 Ry.
For the calculations in the paramagnetic state,
we considered Ba122 in the tetragonal I4/mmm structure (see Fig.\
\ref{fig:smallstructure}), where we used lattice constants $a =
3.9625$ \AA\ and $c = 13.0168$ \AA ~as obtained by Rotter et
al.\cite{m_rotter_08a} for the structure at 297K.
For the undoped SDW state, which is in the orthorhombic Fmmm state,
we used $a = 5.6146$ \AA, $b=5.5742$ \AA\ and $c = 12.9453$ \AA.
To consider a doping $x = 1/16$, we used a supercell of 40 atoms (16 Fe atoms)
and replaced a single Fe by Co.
In the actual material, however, the dopant is not placed preferentially in one plane over another;
all planes are equally likely to be doped. To address this issue, we have additionally done calculations for an
even larger cell of 80 atoms (32 Fe atoms) where two Fe atoms were replaced.
The lattice constants for the doped calculation
were kept the same as for the undoped SDW state. In all the cases reported except the
80 atom cells, the atomic positions were optimized. Where appropriate, we have checked using the 40 atom
cell that the quantity reported for the 80 atom cell does not vary appreciably upon relaxation.

\section{Comparison of the spin-density wave and paramagnetic state}
\begin{figure}[h]
\includegraphics[width=0.5\textwidth, clip = true]{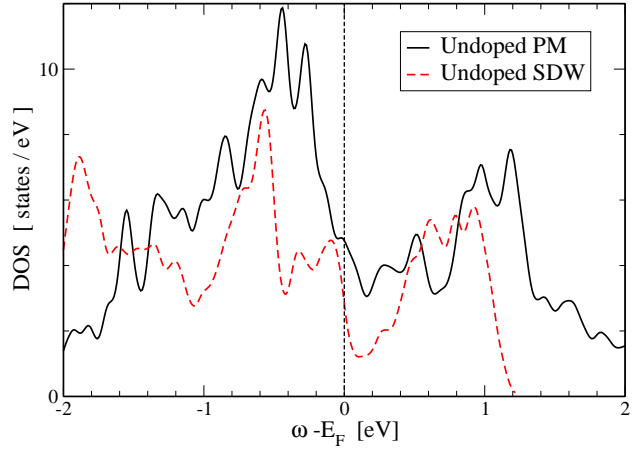}
\caption{DOS for \BaFeAs in the undoped PM and undoped SDW states. The Fermi levels for both systems have been aligned at $0$.}
\label{fig:doscompare}
\end{figure}

The \BaFeAs system has been experimentally observed to support a
spin-density wave (SDW) state, where the spins are aligned along
lines in the (110) direction in the plane, and the planes are
coupled antiferromagnetically\cite{huang_08,zhao_08}. We find the SDW,
orthorhombic state to have a lower energy by 0.55 eV per primitive unit cell.
Fig. \ref{fig:doscompare} shows the density of states for
undoped \BaFeAs in the SDW and PM states. There is a clear
difference in peak structure, as well as a drop in the DOS at the
Fermi level from the PM to the undoped SDW state. As discussed by
Zhang et al.\cite{y_zhang_08}, the drop in the DOS may be caused
by the removal of the incipient magnetic instability associated
with the peak in the PM DOS and the possibility of Fermi surface
nesting.

\begin{figure}[h]
\includegraphics[width=0.5\textwidth, clip = true]{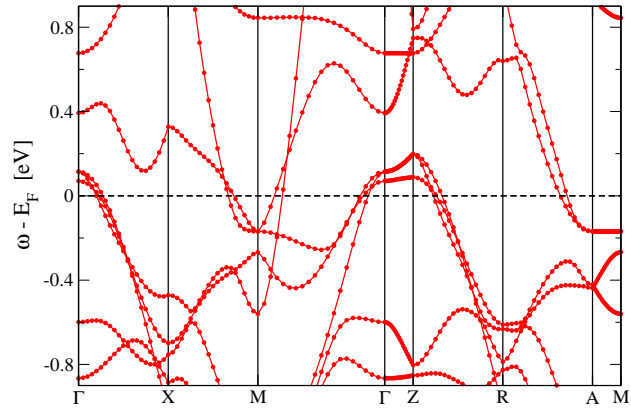}
\caption{Undoped PM band structure along high-symmetry lines}
\label{fig:undoped_pm_bands}
\end{figure}

\begin{figure}[h]
\includegraphics[width=0.5\textwidth, clip = true]{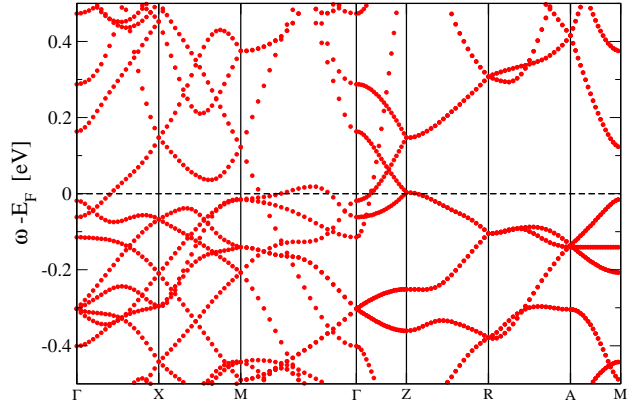}
\caption{Undoped SDW band structure along high-symmetry lines. Due
to symmetry in up- and down-spin, the individual spin states are
degenerate and thus only one is shown.}
\label{fig:undoped_sdw_bands}
\end{figure}

To further study the difference in electronic structure, we have calculated the band structure
for both states, which are shown in Figs.
\ref{fig:undoped_pm_bands} and \ref{fig:undoped_sdw_bands}. For
easier comparison, we have used the tetragonal cell and its
high-symmetry points in all band structure plots. As can be seen
from the figures, the band structure of the SDW state is entirely
different from that of the PM state. The PM band structure, in
agreement with that reported by other
studies\cite{a_sefat_08,a_leithe-jasper_08,a_sefat_08-2,d_kasinathan_09},
has hole pockets around the $\Gamma$ point and electron pockets
around M. Since these pockets are of similar size, there is a
large potential for nesting. In the SDW state, on the other hand
the  nesting has entirely disappeared. Furthermore, there are
Fermi surface crossings from R to A in the PM state which are not
found in the SDW state.

\section{Effect of Co doping}
\begin{figure}[ht]
\includegraphics[width=\textwidth]{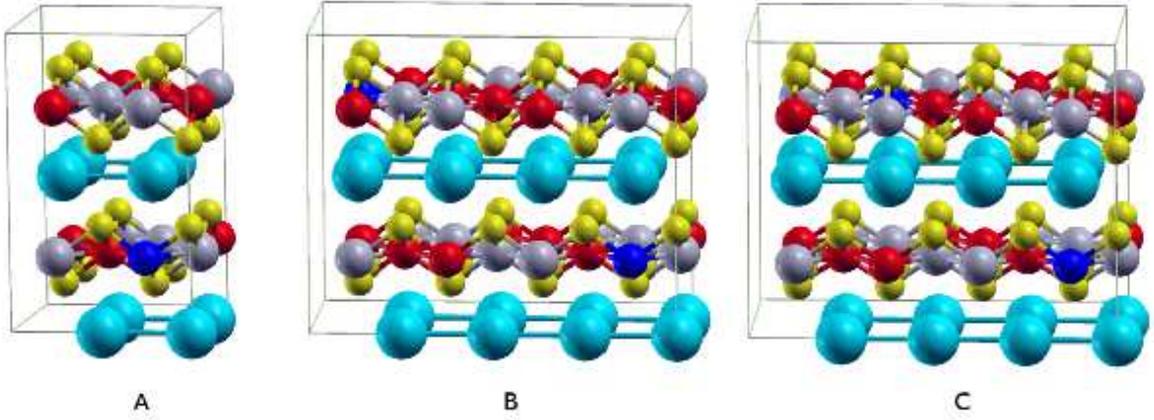}
\caption{(Color online) Configurations of \BaFeCoAs for x=\sixteenth, in the SDW state. (A) 40-atom unit cell with a single Co dopant
(B) 80-atom unit cell with two Co dopants of opposite spin
(C) 80-atom unit cell with two Co dopants of same spin. Ba atoms are light blue, As atoms are yellow,
gray and red balls denote Fe atoms of up and down spin, and the Co dopants are dark blue. Note that configurations B and C have
2 Co dopants each, while maintaining the same concentration.}
\label{fig:structure}
\end{figure}

To investigate the effect of cobalt doping, we used several sets
of atomic configurations, with one of every 16 iron atoms replaced
by a Co atom, yielding a concentration of 6.25\% and an empirical
formula of
Ba(Fe$_{\textnormal{0.9375}}$Co$_{\textnormal{0.0625}}$)$_{\textnormal{2}}$As$_{\textnormal{2}}$,
which is well within the experimentally determined range for the
SDW state as reported by several
studies\cite{s_takeshita_08,x_wang_08,j_chu_08,f_ning_08,n_ni_08,r_gordon_08}.
The configurations considered are shown in Fig.
\ref{fig:structure}. We shall refer to the configuration with a
single Co dopant as configuration A, and to those with two Co
dopants as configuration B for unlike-spin dopants, and
configuration C for like-spin dopants. For configuration A, the
ions were allowed to relax (all within the SDW state) after Co
substitution. The majority of displacements occured near the
dopant site, where the Co atom pushes away the Fe atoms that lie
along the line of collinear spins in the SDW state. Furthermore,
the Co atom attracts the nearest As atoms, causing the distance to
decrease by 0.03\AA\ compared to the normal Fe-As distance.

\begin{figure}[ht]
\includegraphics[width=0.5\textwidth, clip = true]{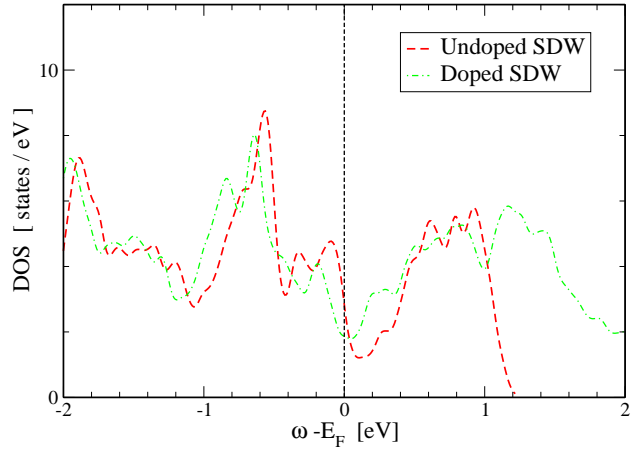}
\caption{DOS for \BaFeCoAs in the undoped and doped configuration
A, SDW states. The Fermi levels for both systems have been aligned
at $0$. } \label{fig:doscompare2}
\end{figure}

Initially limiting considerations to configuration A, we observe that Co
doping induces a small shift in the overall density of
states near the Fermi level (see Fig. \ref{fig:doscompare2}). When the Fermi levels are
aligned, the doped and undoped DOS are similar in shape, suggesting that a
rigid band approach to doping the system is qualitatively valid. However, a closer look will reveal that there are other effects
not captured by a rigid band shift.

\begin{figure}[ht]
\includegraphics[width=0.5\textwidth, clip = true]{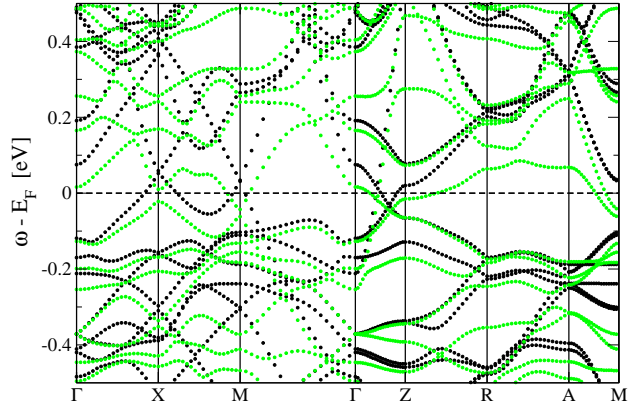}
\caption{(Color online) Doped SDW band structure along high-symmetry lines. Black (green) indicates the majority (minority) spin.}
\label{fig:sdw_bands}
\end{figure}
To investigate the effect of the Co dopant on the spin structure
of the system, we have calculated the band structure along
high-symmetry directions, as shown in Fig. \ref{fig:sdw_bands}.
It is immediately obvious from the picture that the degeneracy
between up and down electrons is broken. The presence of the Co
dopant changes the net spin of the system from $0$ to $0.46\
\mu_B$/cell. Cobalt not only renormalizes the moment on the site
where it it substitutes for Fe, however, but is found to have a
strong non-local effect, decreasing the magnetization over the
entire plane. Some of the polarization is spread from the Co site
to the the other Fe sites in the doped plane (0.05 \textendash
0.10 $\mu_B$ per Fe ion).   Additionally, the presence of the Co
ion enhances the polarization in the undoped layer (in
configuration A), increasing the absolute spin per ion by
approximately 0.05 $\mu_B$.

\begin{figure}[ht]
    \includegraphics[width=0.5\textwidth, clip = true]{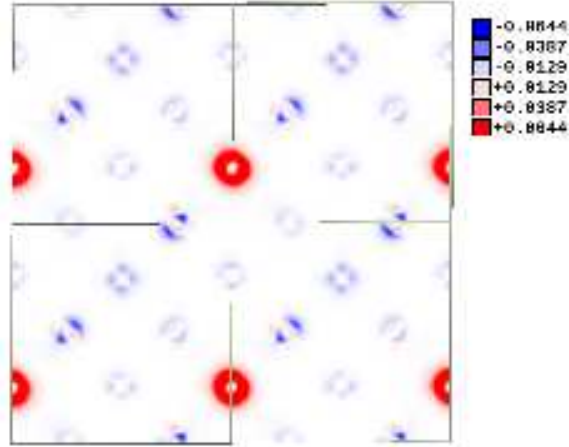}
    \caption{(Color online) Local spin polarization in the dopant plane. A
relatively large polarization is induced around the dopant site, in addition
to a change in polarization on nearby Fe sites of like spin.}
    \label{fig:magnetization}
\end{figure}

\begin{figure}[ht]
\includegraphics[width=0.5\textwidth, clip = true]{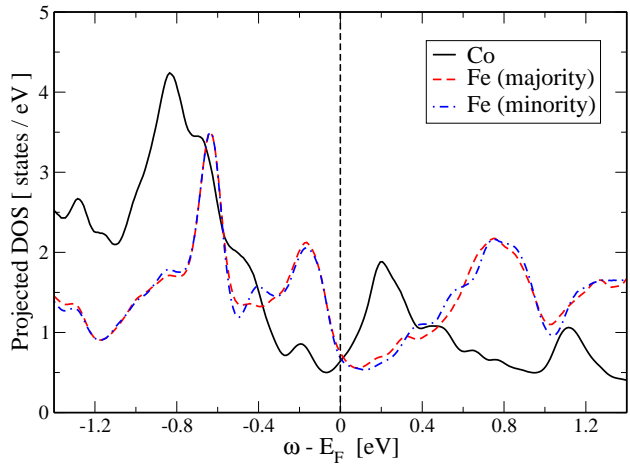}
\caption{(Color online) Projected density of states (PDOS) for
atomic species Co (dashed) and Fe (solid).  The Fe states belong
to atoms in the unit cell located as far as possible from the Co.}
\label{fig:PDOS}
\end{figure}

 The electron doping
also raises the overall Fermi energy, but since the spin
degeneracy is broken, does so differently for the majority versus
the minority spin. Furthermore, the number and character of the
bands crossing the Fermi surfaces changes greatly. After doping, a
minority spin band crosses the Fermi level from $A$ to $M$, and a
number of crossings have appeared from $X$ to $M$.  It is
interesting to ask which of the states remaining near the Fermi
level have Co character, which might enable imaging of the Co
atoms on the surface of this system by STM, as was done
successfully with O dopants in the high-$T_c$ superconductor
Bi$_2$Sr$_2$CaCu$_2$O$_{8+x}$\cite{mcelroy_05}.  This can be
accomplished by plotting the partial density of states (PDOS)
projected onto atomic orbitals of a given species.  In Fig.
\ref{fig:PDOS}, we compare the Co density of states in the doped
system of configuration A with that associated with Fe states
located as far as possible from the Co impurities.  We see very
clear resonances in the Co signal at -800 and  +200 meV.  The
latter energy agrees quantitatively with the bias used to image Co
atoms on the surface of lightly Co-doped Ca-122 in STM
recently\cite{davis_09}.  Note that the present calculation also
predicts that the Co should be visible as minima in the local
tunnelling current relative to nearby Fe at -200 and +800 meV.

In addition to changes in the overall electronic structure of the
system, we can consider the effective local potential due to Co,
as would be used in a tight-binding Hamiltonian.  For a discussion
of this quantity and its calculation, see Ref.
\onlinecite{wang_05}. To observe only the first-order effect, we
calculated the total potential for a doped system where we did not
allow relaxation, and subtracted the total potential for an doped
system.
\begin{figure}[ht]
\includegraphics[width=\textwidth, clip = true]{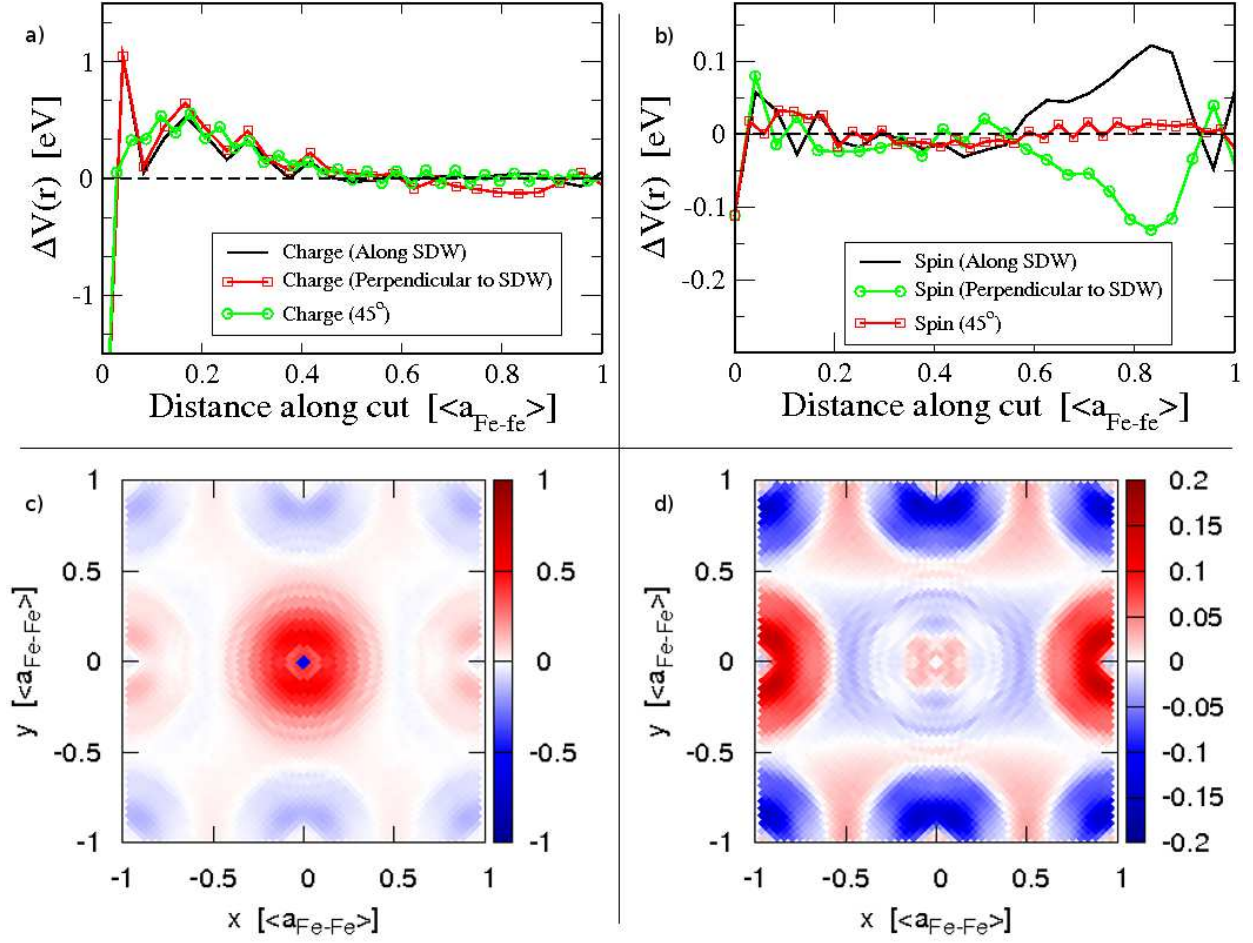}
\caption{(Color online) Left: Cuts of the potential change in the
charge (top) and spin (bottom) channel potentials upon Co doping.
Right: Cuts of the Fourier transformed potential change.}
\label{fig:co-potential}
\end{figure}
Fig. \ref{fig:co-potential} shows a number of line cuts of the
potential change in the two-dimensional Fe plane, scaled to the
average Fe-Fe distance $\langle a_{Fe-Fe}\rangle = 2.7972$\AA
(note we use here the coordinate axes (x,y) aligned along the
nearest neighbor Fe-Fe bonds). The directions are oriented
according to the lines of collinear spins of the SDW state. The
nonmagnetic potential variation is localized to roughly 1 \AA\
from the dopant center, or about 0.25 of  an average Fe-Fe
distance. The oscillations visible in the figure
 occur near the core and originate from the difference in electron
shell structure between Co and Fe, i.e. the larger Z for Co causes
a difference in the radial shell structure as compared to Fe. In
the spin channel, the changes in potential occur mainly near the
next iron site, indicating that there is a strong non-local effect
in the magnetic behavior of the system.  In {\it both} cases, it
is clear that the collinear SDW state couples very strongly to the
Co and drives a large anisotropy in the effective potential.

In models of correlated Fermi systems, a nonmagnetic impurity can
induce magnetization on neighboring sites through its perturbation
of the magnetic correlations in the host material; such effects
are well known in quantum spin systems and high-$T_c$
cuprates\cite{alloul_09}, and can be qualitatively captured in
mean field theories. The current calculation describes the
intrinsic magnetic effects due to the Co ion itself, as well as
induced magnetic effects of this type in the surrounding Fe system
at the DFT level, and should be a good approximation to the actual
potential due to the intermediate strength correlations in these
systems. In the simplest cases, one expects such induced magnetic
effects due to correlations to decay away from the impurity site
with a length scale corresponding to the magnetic correlation
length in the pure host. Here  we note that, in contrast to the
charge sector potential, we can detect no significant decay of the
magnetic potential over the size of our unit cell at all,
consistent with the long range order present in the ground state.
The long-range effect of this potential may be responsible for the
rapid destruction of magnetic order by Co observed in this
system\cite{canfield_09}.

Calculations discussed thus far refer to the effect of a Co on the
system in its SDW ground state.  In principle, it is interesting
to discuss the same questions in a host PM state. However, we find
that the introduction of a Co in the PM state always induces   a
SDW state spanning our entire unit cell, in both the orthorhombic
and tetragonal lattices, and for a number of initial conditions.
This fact, coupled with the discussion above of the strong
nonlocal component to the potential in the spin channel  implies
that in the PM state the scale on which the dopant affects the
magnetic behavior is longer than our unit cell size.



In a simple two-band phenomenological picture for the Fe plane
used frequently in the literature, the two sheets of the Fermi
surface are connected by a vector $(\pi/a,0)$ or $(0,\pi/a)$ in
momentum-space.  Elastic electronic scattering processes which
connect these sheets are referred to as interband, and are
considered particularly important because they will break Cooper
pairs in superconducting states like the currently popular
``$s_{\pm}$" state proposed by Mazin et al. as a candidate for the
ground state in the Fe-pnictide systems\cite{mazin_08}.  Since the
spin potential shown in Fig. \ref{fig:co-potential} is strongly
peaked at a distance close to one Fe-Fe distance, a Fourier
transform will be strongly peaked at precisely the interband
scattering wavevectors.  One might be tempted to deduce a strong
interband scattering component for use in simple phenomenological
models of Co as a localized elastic scatterer. This is incorrect,
however, since the wave vectors must be considered in the magnetic
Brillouin zone where the ``interband" $\bf q$ are equivalent to
zero in the system with long range commensurate order.


We would nevertheless like to use the information from the DFT
calculation to provide an input to phenomenological models, and
therefore project the potential onto local atomic states, with the
caveat that the following analysis is only valid in the system
with long range collinear magnetic order.   To be able to map the
potential to an on-site energy within a tight-binding model, the
quantity needed is the matrix element of the potential in the
appropriate Fe 3d orbital (which we denote by $U^m_c$),

\begin{align}
 U^m_c(\vec{R}) = \sum_{\sigma} \langle\ \phi_{l=2}^m(\vec{r}-\vec{R}) | \Delta V_\sigma(\vec{r}) |
 \phi_{l=2}^m(\vec{r}-\vec{R})\rangle,
\end{align}
where we have used the radial atomic wavefunction from the pseudopotential.
Table \ref{table:potentials} lists projections of the matrix elements for various sites near the impurity.
Note that there are two neighbors of each type, and we have reported the potential averaged on the two (although
they do not deviate appreciably from the average).

In addition to scattering in the charge channel, it is also useful
to calculate the scattering in the magnetic channel. To do so, we
calculate the difference between the spin up and spin down
scattering potentials, projected onto the same iron 3d orbital:

\begin{align}
 U^m_s(\vec{R}) = \sum_{\sigma} \langle\ \phi_{l=2}^m(\vec{r}-\vec{R}) | \sigma\Delta V_\sigma(\vec{r})
  | \phi_{l=2}^m(\vec{r}-\vec{R})\rangle,
\end{align}
where $\sigma$ is either $\half$ or $-\half$. The values found are
shown  in Table \ref{table:potentials}. Note that the sign of the
scattering in the spin channel will depend on which
spin-sublattice the Co dopant is located. As can be seen from the
Table, as the cobalt ion changes the potential on the impurity
site, it also induces significant nearest-neighbor potentials. The
neighboring Fe sites with the same spin state slightly repel both
electrons and spin, and the sites in the perpendicular direction
attract both.

\begin{center}
\begin{table}[h]
\begin{tabular*}{\textwidth}{@{\extracolsep{\fill}}c|c|cc|cc|cc}
\hline\hline
      &          & \multicolumn{2}{c|}{Impurity site} & \multicolumn{2}{c|}{Same-spin neighbor} & \multicolumn{2}{c}{Opposite-spin neighbor} \\
 $m$  & Orbitals & $U_c^m$ & $U_s^m$ & $U_c^m$ & $U_s^m$  & $U_c^m$ & $U_s^m$\\
\hline
\hline
 0       & $d_{z^2}$        & 1.29 & -0.331 & 0.572 & 0.709 & -0.634 & -0.639\\
 $\pm 1$ & $d_{xz}, d_{yz}$ & 1.52 & -0.378 & 0.672 & 0.848 & -0.805 & -0.733\\
 $\pm 2$ & $d_{x^2-y^2}, d_{xy}$& 1.87 & -0.495 & 0.882 & 1.099 & -0.956 & -0.965
 \end{tabular*}
\caption{Projections of the scattering potential onto $d$ orbitals on both the cobalt dopant site and the nearest neighbor Fe sites,
in both the charge and spin channels. All values are in eV.}
\label{table:potentials}
\end{table}
\end{center}

\begin{figure}[ht]
\includegraphics[width=0.7\textwidth, clip = true]{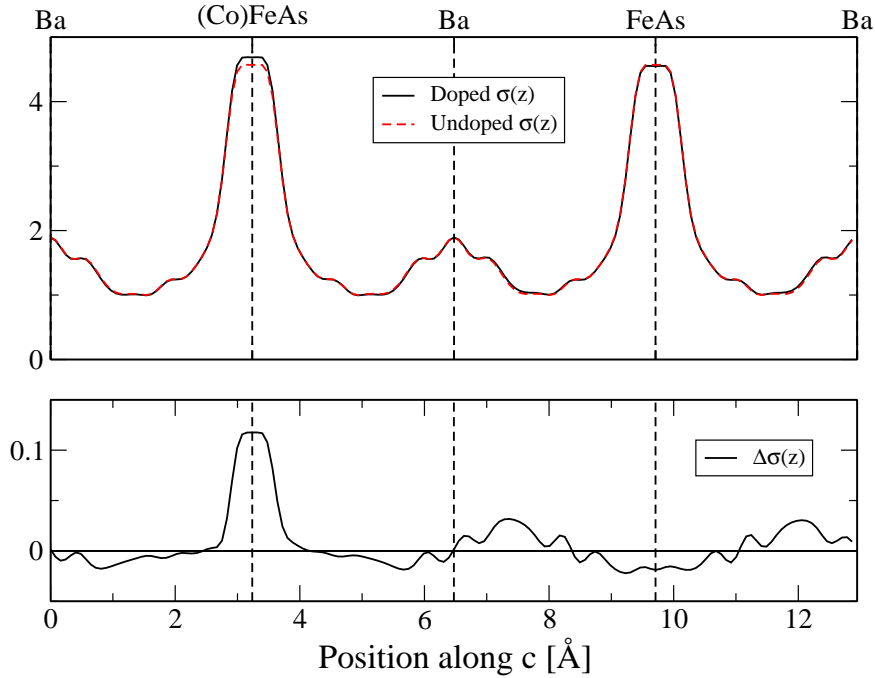}
\caption{(Color online) Top: linear integrated charge density for the doped and undoped systems, in number of electrons. Bottom: difference between doped and undoped
linear integrated charge densities}
\label{fig:charge-z}
\end{figure}

We now consider the process by which Co dopes the FeAs plane. Fig.
\ref{fig:charge-z} shows the linear integrated planar charge
density $\sigma(z) = \int dx\ dy\ n(\vec{r})$ As can be seen from
the figure, there is very little doping outside of a 2 \AA\ range
around the plane containing the Co dopant. There is, however, some
shift of charge around the undoped plane, due to the relocation of
the As atoms. However, integration around the planes finds that
the net charge transferred to the undoped plane is zero, i.e. the
whole doped electron remains in the dopant plane. We have verified
 that the doped electron is delocalized in this plane
over the extent of our current unit cell (not shown).

\begin{figure}
\centering
    \includegraphics[width=0.8\textwidth]{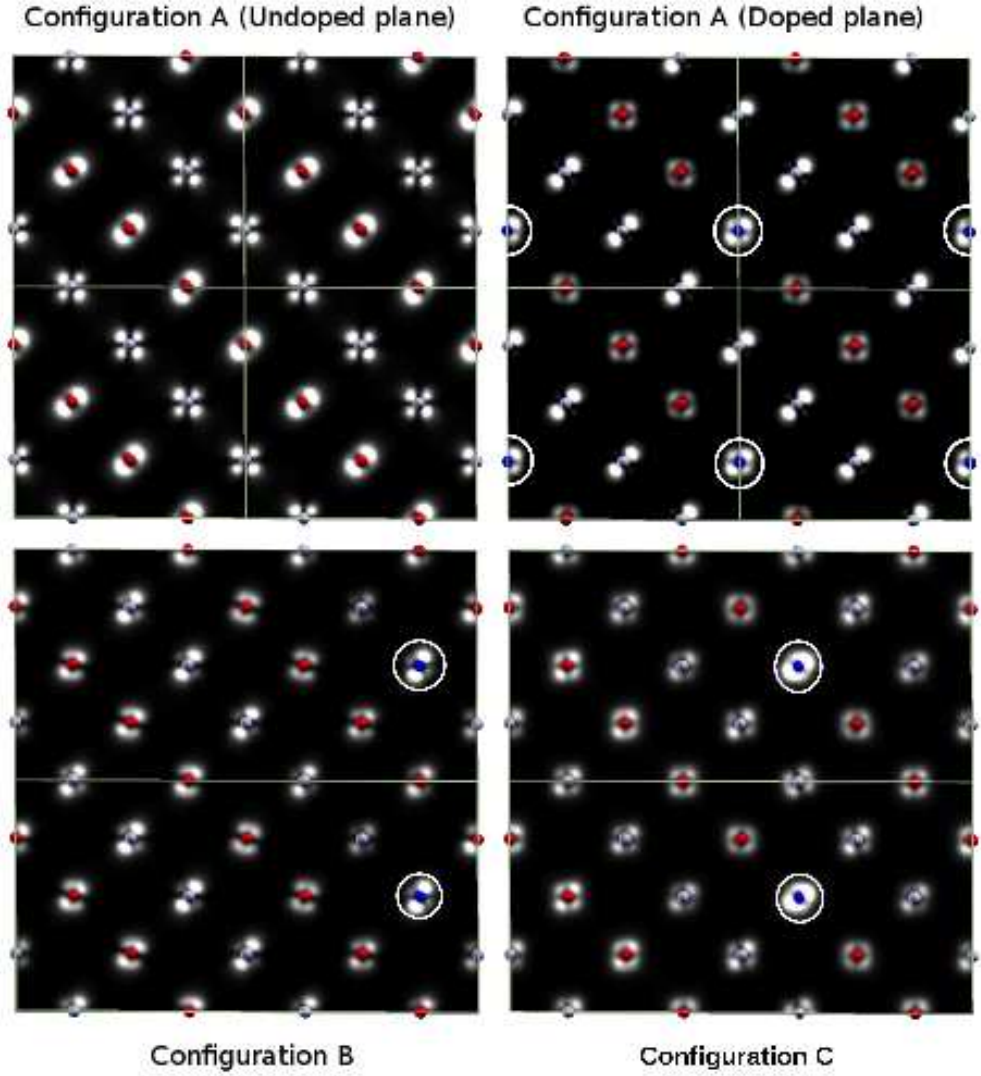}
    \caption{(Color online) Plane cuts of the local Fermi level DOS (see Eq. \ref{eq:ldos}).
Red (gray) balls indicate Fe ion in the spin down (up) state.
The Co dopant is indicated blue, has been circled where visible. Four unit cells are shown for configuration A, and two unit cells are shown for
configurations B and C.}
\label{fig:doped-ldos}
\end{figure}

The effective potential due to the dopant considered above will
affect the surrounding density of states near the Co; in
particular it may be interesting to consider the apparently
long-range effect of the magnetic potential on states near the
Fermi level.   To this end we calculate the local density of
states, integrated over a small range around the Fermi energy:
\begin{align}
 \nu(\vec{r}) = \sum_\sigma \int_{E_F - \delta}^{E_F + \delta} d\omega
 \rho_\sigma(\vec{r},\omega),
\label{eq:ldos}
\end{align}
where we have chosen $\delta = 0.01 eV$.
Fig. \ref{fig:doped-ldos}A shows a cut of $\nu(\vec{r}$) through the undoped and doped Fe
plane for configuration A.
There is a clear modulation of the density of states in both
planes; the modulation is commensurate with the spin-density wave,
i.e. $\nu(\vec{r}$) is increased along the lines of majority spin,
and suppressed along those of minority spin. Thus, the Co dopant
in configuration A produces effects in the local DOS of the
neighboring plane, even though it does not dope this plane.
Figs. \ref{fig:doped-ldos}B and \ref{fig:doped-ldos}C show cuts
of $\nu(\vec{r}$) through one of the planes of configurations B
and C. The atomic positions for configurations B and C are not
optimized; we have, however, compared $\nu(\vec{r})$ for optimized
and non-optimized configuration A, and find that there is no
appreciable qualitative difference. Since there are two Co
dopants, depending on their location relative to the magnetic
structure, there is the possibility of them being in the same- or
opposite-spin states. Additionally, there will be interference
effects based on their relative location, but we do not address
this issue  here. We have placed them as far apart as possible,
and will restrict ourselves to considering the first-order effects
on the local DOS. As can be seen from the figure, the clear
stripes seen in the undoped plane of configuration A are not as
pronounced in either configuration B or C. There is still some
difference between the lines of up- and down-spin, but the
contrast is markedly smaller. Since these effects occur in the
total (spin-summed) local DOS, they may be visible in standard STM
experiments. Even though the effect in configurations B and C is
much smaller, the dopant distribution of the samples are not
always uniform, as found in some NMR experiments\cite{f_ning_08}.
So, all three configurations considered above are possibly visible
experimentally.  The importance of this result is to note that the
long-range magnetic perturbation of a Co dopant on the spin system
should have a significant effect modulating the electronic
structure of the FeAs plane; depending on the nature of the Co-Co
interference effects largely neglected here, these modulations may
occur as well on longer length scales than our current unit cell
size, and may be responsible for the 8-Fe lattice spacing
incoherent modulations reported in Ref. \onlinecite{davis_09}.

\section{Three-dimensionality of B\lowercase{a}F\lowercase{e$_2$}A\lowercase{s$_2$} }

\begin{figure}[h]
\includegraphics[width=\textwidth, clip = true]{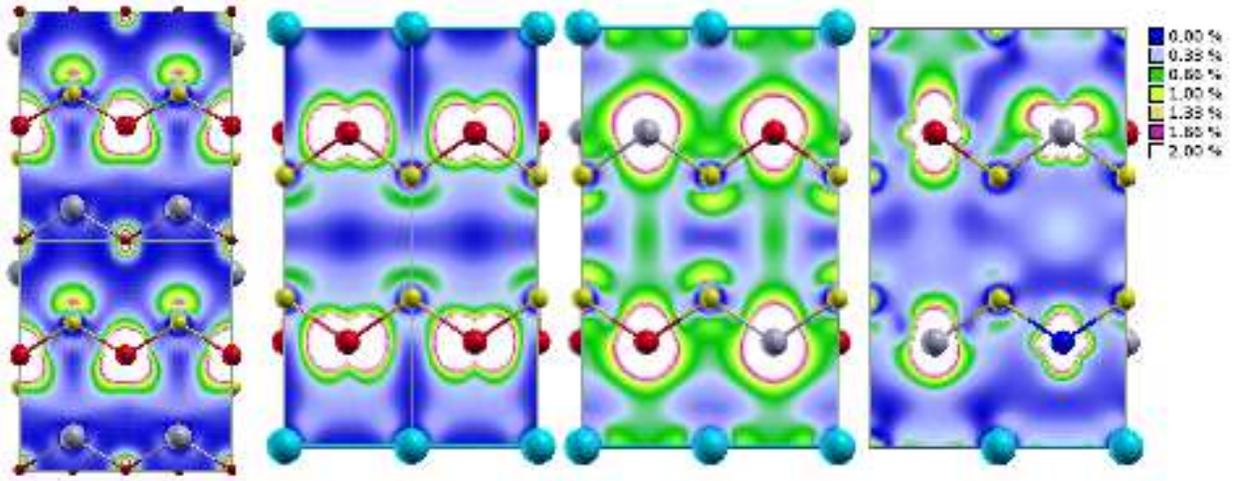}
\caption{(Color online) Cut of the local Fermi level DOS (see Eq.\ \ref{eq:ldos}) through the vertical plane FeAs plane for \LOFA\ in the PM state,
\BaFeAs in the PM state \BaFeAs in the SDW state and \BaFeCoAs (x=\sixteenth) in configuration A.
Colors are scaled from 0\% to 2\% of the maximum local DOS in the horizontal FeAs plane. For \BaFeAs (undoped and in configuration A),
red (gray) balls indicate spin down (up) Fe atoms.}
\label{fig:ldos-vertcut}
\end{figure}

As mentioned above, critical field, transport and London
penetration depth data appear to indicate that the 122 systems are
considerably less two-dimensional than their 1111 counterparts.
\cite{r_prozorov_09}
In Fig.
\ref{fig:ldos-vertcut} we show the density of states integrated
over a small range around the Fermi level ($\nu(\vec{r})$) along
the plane perpendicular to (100). To illustrate the density of
states present in between the FeAs layers, we have scaled the
plots by the maximum found in the FeAs layer in each case. It is
clear from the figure that \BaFeAs has a larger Fermi level
density of states available between the FeAs layers than \LOFA, which is not
immediately visible from the band structure. This is true for both
the PM and SDW states, suggesting
that it is lattice structure rather than the magnetic state
that is the cause for the difference.
The distance between FeAs layers is roughly 3\AA\ less in \BaFeAs
than it is in \LOFA, accounting for the larger c-axis
dispersion seen in the 122-structure materials\cite{d_singh_08} as
compared to the 1111-type materials, whose Fermi surfaces are
generally considered to be cylindrical. Furthermore, in \LOFA, the
As $z$ position is in phase from layer to layer, whereas in
\BaFeAs it is out of phase \textemdash the interlayer As atom
pairs are alternately closer and farther (see Fig.
\ref{fig:ldos-vertcut}). This explains the presence of additional
states between the layers.
Upon cobalt substitution, the density of states available in between
the planes decreases (rightmost panel of Fig. \ref{fig:ldos-vertcut}),
suggesting the Co doping increases the anisotropy.
So far, the theoretical models
for the pnictide superconductors have been limited to two dimensions. These results, as well as the experimental
work,\cite{r_prozorov_09,m_tanatar_08,n_ni_08} suggest that a three-dimensional model is needed.

\section{Conclusions}
We have performed an in-depth study of cobalt doping and
anisotropy in the oxygen-free pnictide superconductor \BaFeCoAs
for x=\sixteenth. For this low doping, it is appropriate to
consider the SDW state, which is quite different in nature from
the PM state. Relative to the more commonly discussed PM
electronic structure, in the SDW state the number of bands
crossing the Fermi surface  and the density of states at the Fermi
level are  reduced. Upon Co doping, the density of states at the
Fermi level is shifted slightly, but there are other changes in
local electronic structure which are not consistent with a rigid
band shift. Co was found to only dope the plane where it resides,
and the charge of one electron is spread over the entire plane as
far as can be determined with our unit cell. Furthermore, it
provides both a local and a nonlocal scattering potential, of
roughly 1.5 eV in the charge channel, and 0.3 eV in the spin
channel on the impurity site. The majority of the larger
charge-channel potential is localized in a region of approximately
1\ \AA\ in size in real space.
In addition, however, the Co breaks the spin degeneracy of the SDW
state by introducing a net moment of $\sim 0.5\ \mu_B$/cell, and
creates a significant and highly anisotropic nonlocal magnetic
potential on the neighboring Fe orbitals whose range is larger
than our current unit cell size.  This remarkable effect explains
why Co is so effective in suppressing the SDW transition
temperature.

We have furthermore investigated the effect of Co on the
electronic  states near the Fermi level. The Co states themselves
are responsible for resonances we find at -800 and +200 meV, the
latter of which has already been used by STM\cite{davis_09} to
image Co atoms on the surface of Co-doped Ca-122. An further
intriguing consequence of the long-range nature of the magnetic
potential is that the Co induces a modulation of the local density
of states aligned with the wavevector of the SDW state, which may
also be visible by STM. The determination of the actual range of
the magnetic influence of Co, and the effects of interference of
many Co, which may lead to other longer range modulations of the
system, will be pursued in a later study.


Finally, we have addressed the issue of the low anisotropy,
comparing \LOFA ~and \BaFeAs by examining the local density of
states near the Fermi energy. We find there to be a larger number
of states available between FeAs layers in both the PM and SDW
state of \BaFeAs, partially due to the interlayer As position;
this appears to be the cause for the lower anisotropy observed in
experiment. Co doping appears to reduce the coupling between
planes and therefore enhance the anisotropy.  These results point
to a need for three-dimensional models to properly describe these
systems.

\begin{acknowledgments}
This work was supported by DOE grants DE-FG02-02ER45995 (HPC) and
DE-FG02-05ER46236 (PJH). We thank J.C. Davis, S. Pan, G. Sawatzky,
and R. Valenti for helpful discussions, and  H. Kontani for a
useful suggestion. The authors acknowledge the University of
Florida High-Performance Computing Center for providing
computational resources and support that have contributed to the
research results reported within this paper. URL:
http://hpc.ufl.edu
\end{acknowledgments}
\bibliography{master}
\end{document}